# SchrödingerNet: A Universal Neural Network Solver for The Schrödinger Equation

Yaolong Zhang[1*], Bin Jiang[2*], and Hua Guo[1*]

[1]Department of Chemistry and Chemical Biology, Center for Computational Chemistry, University of New Mexico, Albuquerque, New Mexico 87131, USA

[2]Key Laboratory of Precision and Intelligent Chemistry, Department of Chemical Physics, University of Science and Technology of China, Hefei, Anhui 230026, China

*: corresponding authors, email: ylzhangch@unm.edu, bjiangch@ustc.edu.cn, hguo@unm.edu

## ABSTRACT

Recent advances in machine learning have facilitated numerically accurate solution of the electronic Schrödinger equation (SE) by integrating various neural network (NN)-based wavefunction ansatzes with variational Monte Carlo methods. Nevertheless, such NN-based methods are all based on the Born-Oppenheimer approximation (BOA) and require computationally expensive training for each nuclear configuration. In this work, we propose a novel NN architecture, SchrödingerNet, to solve the full electronic-nuclear SE by defining a loss function designed to equalize local energies across the system. This approach is based on a translationally, rotationally and permutationally symmetry-adapted total wavefunction ansatz that includes both nuclear and electronic coordinates. This strategy not only allows for an efficient and accurate generation of a continuous potential energy surface at any geometry within the well-sampled nuclear configuration space, but also incorporates non-BOA corrections, through a single training process. Comparison with benchmarks of atomic and small molecular systems demonstrates its accuracy and efficiency.

## INTRODUCTION

The Schrödinger equation (SE) is a cornerstone of quantum mechanics. Solution of the stationary SE yields energies and the corresponding wavefunctions, which in principle provide a complete characterization of the system. However, finding an accurate solution to the SE for a many-body system can be extremely challenging. For molecules, it is generally assumed that the nuclear degrees of freedom (DOFs) are decoupled from the electronic ones as the former move much slower than the latter, given their large masses. Within this so-called adiabatic or Born-Oppenheimer approximation (BOA)[1], the electronic SE is solved at fixed nuclear configurations. The nuclear SE is subsequently solved on the adiabatic potential energy surface (PES) formed by the expectation value of the electronic Hamiltonian at different nuclear configurations. Separating the electronic and nuclear motion significantly reduces the complexity of the full SE, leading to the remarkable success of quantum chemistry in past decades[2]. However, the construction of a PES requires repeated electronic structure calculations and their fitting to a continuous function[3]. Further, the BOA is known to introduce significant errors, particularly near electronic degeneracies[4].

Conventional high-level electronic structure approaches, such as the configuration interaction (CI)[5] and coupled-cluster (CC) methods[6], typically use basis functions to reduce the differential equation, namely the electronic SE, to a set of nonlinear eigen-equations. However, they suffer from steep scaling laws and thus are restricted to small molecules. Although truncated or local approximations allow CI and CC methods to handle large systems with hundreds of atoms[7-9], these approaches may suffer from a lack of size-extensivity or struggle to accurately describe strongly correlated systems. In addition, the appropriateness of a particular treatment might be geometry dependent, due to changes of the dominant electronic configuration. For example, the single reference CC method is known to fail spectacularly when a chemical bond is broken[6]. On the other hand, computationally less expensive alternatives, such as the density functional theory (DFT)[10], are amenable to larger molecules and extended systems, but at the expense of reduced accuracy and lack of the exact exchange-correlation functional.

With the advent of machine learning (ML) algorithms, there is a strong desire to develop alternative approaches for solving differential equations, such as the SE, with better scaling laws and faster convergence behavior. Indeed, combining a neural network (NN) based wavefunction ansatz with the variational Monte Carlo (VMC) framework[11], several ML methods have been proposed for solving the electronic[12-32] as well as the nuclear SEs[33-37]. The common strategy is to take advantage of the variational principle to minimize the expectation value of the Hamiltonian within the MC framework. These methods have shown promise in yielding more accurate results than the traditional "gold-standard" CCSD(T) (CC with singles, doubles, and perturbative triples) on various systems with better numerical scaling than conventional high-level ab initio methods[13-15]. However, most of these NN-based electronic SE solvers ignore the non-BOA correction and require repeated training for each nuclear geometry, resulting in high computational costs in constructing a global PES. Schemes to mitigate this problem have been suggested. For example, a weight-sharing scheme was proposed to reduce the time for each training, akin to using a good trial wavefunction, but separate trainings are still required for different geometries[19]. Gao and Günnemann suggested another solution by parameterizing the orbital functions in the electronic wavefunction ansatz with another NN model that is dependent on nuclear coordinates, which allows them to use a single training to obtain energies for multiple selected nuclear geometries[20-21].

In this work, we propose a novel NN-based approach that directly solves the full SE, without invoking BOA. For the first time, the total wavefunction in both nuclear and electronic coordinates is expressed in terms of NNs and the full Hamiltonian is considered in a loss function that equalizes the local energy (defined below) throughout the coordinate space. A well-designed NN-based model, which is rotationally and translationally invariant, parity distinguishable, and permutational symmetric/anti-symmetric with respect to nuclear and electronic coordinates, is introduced to represent the total wavefunction. The physics-informed symmetry treatment not only reduces the required sampling space, but also facilitates accurate and fast training. A significant advantage of this approach is to accurately determine the total energy for the system through a single training, which not only efficiently yields a continuous PES at any geometry in the well-sampled configuration space, but also includes non-BOA corrections that are difficult to obtain by using conventional electronic structure methods or previous NN-based methods.

## METHODS

### Wavefunction Ansatz

Without loss of generality, the full non-relativistic Hamiltonian of a system with $N_n$ nuclei and $N_e$ electrons can be expressed in atomic units as,

$$\hat{H} = -\sum_{I}^{N_n} \frac{1}{2m_I} \nabla^2_{\mathbf{R}_I} - \frac{1}{2} \sum_{i}^{N_e} \nabla^2_{\mathbf{r}_i} + \sum_{i>j} \frac{1}{\left|\mathbf{r}_i - \mathbf{r}_j\right|} - \sum_{i} \sum_{I} \frac{Z_I}{\left|\mathbf{r}_i - \mathbf{R}_I\right|} + \sum_{I>J} \frac{Z_I Z_J}{\left|\mathbf{R}_I - \mathbf{R}_J\right|}. \quad (1)$$

Here, $\mathbf{r}_i$ and $\mathbf{R}_I$ are Cartesian coordinates of the $i$th electron and $I$th nucleus, with $m_I$ and $Z_I$ as its mass and charge. We choose to work in the Cartesian coordinates to take advantage of the simplicity of the kinetic energy operator, and as discussed below, our wavefunction ansatz effectively eliminates the translation and rotation of the entire system. The eigenenergy of the system ($E$) is associated with its eigenfunction by the time-independent SE,

$$\hat{H}\psi(\mathbf{r},\ \mathbf{R}) = E\psi(\mathbf{r},\ \mathbf{R}), \quad (2)$$

where the total wavefunction $\psi(\mathbf{r},\ \mathbf{R})$ is dependent on both electronic and nuclear positions. Within BOA, the total wavefunction is expanded in terms of electronic and nuclear wavefunctions, with the former being a function of electronic and (parametrically) nuclear coordinates, while the latter depends solely on nuclear coordinates. Instead, we propose here a non-BOA ansatz for the total wavefunction as a simple product of an electronic wavefunction $\psi_e(\mathbf{r},\ \mathbf{R})$ and a nuclear wavefunction $\psi_n(\mathbf{r},\ \mathbf{R})$,

$$\psi(\mathbf{r},\ \mathbf{R}) = \psi_e(\mathbf{r},\ \mathbf{R})\psi_n(\mathbf{r},\ \mathbf{R}), \quad (3)$$

where $\psi_n(\mathbf{r},\ \mathbf{R})$ is also dependent on both electronic and nuclear coordinates. Specifically, $\psi_n(\mathbf{r},\ \mathbf{R})$ is designed to decay smoothly to zero as atoms are far apart or very close to each other,

$$\psi_n\left(\mathbf{r},\ \mathbf{R}\right)=\left(\sum_{m=1}^{N_b} c_m \exp\left[\sum_{I=1}^{N_n}\sum_{I\neq J}^{N_n} -\frac{\left(\alpha_{J,m}\right)^2}{d_{IJ}} - \left(\eta_{J,m}\right)^2 d_{IJ}\right]\right)^2, \tag{4}$$

where $d_{IJ}$ is the inter-nuclear distance. This function is learnable by combining $N_b$ basis functions, for which each combination coefficient $c_m$ is the summation of the NN outputs centered on the nucleus, depending on the positions of the other electrons and the nucleus as shown in Fig. 1. $\alpha_{J,m}$ and $\eta_{Jm}$ are the output of $J$th nucleus-centered NN NN. For simplicity, the nuclear wavefunction is chosen to be positive-definite, thus only suitable for vibrational ground state. Future work on excited states will need modification of the ansatz.

To satisfy the Pauli principle for the fermionic electrons, $\psi_e\left(\mathbf{r},\ \mathbf{R}\right)$ is expressed by a combination of Slater determinants that enforce the anti-symmetry with respect to the permutation of electrons,[14]

$$\psi_e\left(\mathbf{r},\ \mathbf{R}\right)=\sum_{n=1}^{N_{\det}} \det\left[\phi_i^n\left(\mathbf{r}_j;\mathbf{r},\ \mathbf{R}\right)\right]. \tag{5}$$

where $\mathbf{r}_j$ is the corresponding electronic coordinates. $\phi_i^n\left(\mathbf{r}_j;\mathbf{r},\ \mathbf{R}\right)$ represent the $i$th orbital of nth determinant. Different from the Hartree-Fock ansatz where each orbital is only a function of the coordinate of a single electron, importantly, the orbitals are functions of all coordinates, which greatly increase the expressiveness of the wavefunction and can encompass all electron correlations in principle. Each orbital function in the Slater determinant consists of three parts,

$$\phi_i^n\left(\mathbf{r}_j;\mathbf{r},\ \mathbf{R}\right)=\Lambda_{ij}^n(\mathbf{r}_j;\mathbf{r},\ \mathbf{R})\Omega^n(\mathbf{r}_j;\ \mathbf{R})\Theta(\mathbf{r},\ \mathbf{R}). \tag{6}$$

To introduce the correct symmetry, the first part $\Lambda_{ij}^n(\mathbf{r}_j;\mathbf{r},\ \mathbf{R})$ is represented by the $j$th electron-centered NN depending on all other electronic and nuclear positions, thus encompassing all electron correlations. The second part is a sum of exponential functions of the electron-nuclear distance, ensuring that the wavefunction smoothly decays to zero when this electron j is far from all nuclei[14-15],

$$\Omega^n\left(\mathbf{r}_j;\ \mathbf{R}\right)=\sum_I^{N_n}\exp\left(-\left|\alpha_{jI}d_{jI}\right|\right). \tag{7}$$

Here, $d_{jI}$ is the distance between the $j$th electron and $I$th nucleus, $\alpha_{jI}$ is an output of the element-embedding NN. The third part introduces the cusp condition[11-12, 38], which forces the trial wavefunction to have the correct singular behavior when any two particles are close to each other,

$$\Theta\left(\mathbf{r},\ \mathbf{R}\right)=\sum_{i>j}^{N_e}\exp\left(\frac{cd_{ij}}{1+d_{ij}}\right)+\sum_I^{N_n}\sum_i^{N_e}\exp\left(\frac{-Z_I d_{iI}}{1+d_{iI}}\right). \tag{8}$$

Here, $c$ is 0.5 or 0.25 for a pair of electrons with the same or different spins[11]. Note that NN-based wavefunctions need be either O(3) symmetric or antisymmetric. Therefore, the descriptors must be parity distinguishable[39]. More details on our NN architecture can be found below.

Most existing NN approaches for solving SEs have resorted to the VMC formulation[33-36]. However, the VMC approach necessitates integration over the coordinate space, namely the expectation value of the Hamiltonian, which is numerically demanding. Although the quadrature can be replaced by MC sampling, a very large number of sampled points might still be needed to reach the desired accuracy of the variation. Moreover, this method tends to underrepresent the configuration space with low probabilities, as these regions contribute marginally to the overall integral. It is particularly problematic for the overall nuclear-electronic distribution, where nuclear configurations far from equilibrium can lead to very poor performance (unstable energy prediction) in those regions[11].

Instead, we choose to enforce the equalization of the local energy everywhere throughout the electronic and nuclear configuration spaces by using the following loss function,

$$\mathcal{L} = \sum_i \left( E_L(\mathbf{r}_i, \mathbf{R}_i) - E \right)^2, \tag{9}$$

where $i$ denotes the points chosen for the optimization and the local energy is defined by[14],

$$E_L(\mathbf{r}, \mathbf{R}) = \frac{\hat{H}\psi(\mathbf{r}, \mathbf{R})}{\psi(\mathbf{r}, \mathbf{R})}. \tag{10}$$

This local energy-based loss function guides NNs to find an optimal eigenfunction that satisfies the SE and assigns an identical local energy for every single configuration, without the need for integration. In the Cartesian coordinate system, the action of the potential energy operator is realized by simple multiplication, while the kinetic energy operator (the Laplacian) can be evaluated by Forward Laplacian[29, 40] based on the JAX framework. In principle, an exact wavefunction of the system would make the defined loss function in Eq. (9) vanish in all configurations. In practice, however, $\mathcal{L}$ is minimized at a finite number of representative configurations. Since this approach is not variational, we can freely choose a suitable sampling strategy. In addition, it allows us to employ mini-batch optimization, which can make use of parallel computing with multiple modern graph processor units (GPUs) and a massive dataset. Moreover, minimizing this loss function does not require the sampled data to be subject to a specific distribution ( $\sim|\psi(\mathbf{r},\ \mathbf{R})|^2$ ) as in VMC, as long as they adequately cover the desired configurational space.

Numerically, the sampling of the configuration space plays a critical role. Our sampling starts with a set of nuclear configurations in a physically relevant region and some randomly sampled electronic configurations following a Gaussian distribution around each nuclear configuration. An initial wavefunction is randomly chosen and a few MC steps are then run based on the trial wavefunction. Gradient descent optimization is performed until $\mathcal{L}$ is smaller than 90% of the training error of the previous epoch. Subsequently, another few MC steps are executed to identify and add additional configurations, which are selected from those with errors being 2~6 times that of the previous step, to the training dataset. The wavefunction is updated on the new dataset and the next cycle of MC sampling is repeated until convergence. Using this iterative strategy, which required approximately 1-4 hours on an A800 GPU, around 20,000 configurations were sampled to describe each of the systems investigated in this work.

It should be noted that similar concepts have been proposed for solving the electronic or nuclear SE based on the eigenfunction definition[26, 41], but these approaches typically employed a loss function that explicitly depends on the wavefunction, e.g., $\mathcal{L}=\sum_i\left(H\psi(\mathbf{r}_i,\mathbf{R}_i)-E\psi(\mathbf{r}_i,\mathbf{R}_i)\right)^2+\mathcal{L}_{reg}$ . Such methodology has two significant disadvantages. First, this loss function includes a trivial and invalid solution where the wavefunction is zero everywhere. Second, this loss function tends to underestimate the significance of regions where the wavefunction has small absolute values. In contrast, our loss function is solely based on the local energy, which does not explicitly depend on the wavefunction and describes regions with small wavefunction values with comparable precision. This is found to be extremely important for solving the full nuclear-electronic SE. Another similar concept is the collocation method mostly used for solving nuclear SE, even using Cartesian coordinates[42]. This method has been also applied to solving the electronic Schrödinger equation, but limited to systems involving a single electron[43].

**NN framework**

The SchrödingerNet architecture is illustrated in Fig. 1. The key parameters of the nuclear and electronic wavefunctions are represented by equivariant message passing neural networks framework[39], which is not only rotationally invariant but also parity distinguishable. To make this NN framework amenable to both electrons and nuclei, we introduce an embedding NN that maps both atomic numbers of nuclei and spin quantum numbers of electrons to an array of parameters that are used subsequently,

$$\{\boldsymbol{\alpha}, \mathbf{c}, ...\} = f\left(Z_i Z_j,\ Z_i + Z_j\right), \tag{11}$$

where $f$ is an embedding NN whose architecture is the NN module as shown in Fig. 1(c), whose structure is same as that described in our previous work[44]. $Z_i/Z_j$ is either the atomic number for a nucleus or a fractional number with 0.5 indicating a spin-up electron and -0.5 a spin-down electron. This design treats electrons with different spins as completely different particles and is compatible with the unrestricted Hartree-Fock formalism adopted by Eq. (5).

In order to better describe the nucleus-centered and electron-centered environment, we start from a series of contracted Gaussian-type orbitals (CGTOs) in spherical representation. Specifically, for the $i$th nuclear or electronic center,

$$\chi_{lmk}\left(\vec{\mathbf{r}}_{ij}\right) = Y_{lm}(\vec{\mathbf{r}}_{ij})\sum_{n}^{N_\chi} \omega_{n,k}^{t+1} \exp\left[-\beta_n \left(d_{ij} - d_n\right)^2\right], \tag{12}$$

where $Y_{lm}(\vec{\mathbf{r}}_{ij})$ are spherical harmonics of a vector pointing from the $j$th neighboring particle to the $i$th center, both can be either an electron or nucleus, , $d_{ij}$ is the corresponding distance between them, $\beta_n$ and $d_n$ are learnable parameters as the output of the embedding NN ($n$=1, 2, …, $N_\chi$ ). $\omega_{n,k}^{t+1}$ is the $n$th contraction weight corresponding to the $k$th CGTO[44]. We emphasize that the CGTOs used in our work are not real quantum-chemical orbitals but serve as equivariant features to construct a general representation for atomic structures, as discussed in our previous work[44-45]. This approach is analogous to strategies commonly used in constructing machine-learning potential energy surfaces[3, 46]. A linear combination of these CGTOs over all neighboring particles results in an equivariant feature,

$$\varphi_{i,lmk}^{0} = \sum_{j\neq i}^{N} c_j^0 \chi_{lmk}\left(\vec{\mathbf{r}}_{ij}\right). \tag{13}$$

Then, in the interaction module as shown in Fig. 1, we can generate the equivariant features by the tensor product of the CGTO with the corresponding $\varphi_{j,l_f,\mathrm{m}_f,k}^{t}$ with $l$ ranging from 0~$L$ ($L$ is a hyper-parameter listed in Table I),

$$\varphi_{i,lmk}^{t+1} = \sum_{j \neq i}^{N} c_j^t \sum_{m_f, m_i} C_{l_i, \mathrm{m}_i, l_f, \mathrm{m}_f} \varpi_{l_i, l_f, l} \chi_{l_i, \mathrm{m}_i, k} \left( \vec{\mathbf{r}}_{ij} \right) \varphi_{j, l_f, \mathrm{m}_f, k}^t , \tag{14}$$

whose square yields an invariant density feature,

$$\rho_{i,lk}^{t+1} = \sum_{m=-l}^{l} \left( \varphi_{i,lmk}^{t+1} \right)^2 , \tag{15}$$

where $C_{l_i, \mathrm{m}_i, l_f, \mathrm{m}_f}$ is the corresponding CG coefficients and $\varpi_{l_i, m_i, l_f, m_f}$ is the output of the embedded NN. Note that Eq. (14) is a message passing form, where the scalar coefficient $c_j^t$ and the equivariant feature $\varphi_{i,lmn}^{t+1}$ are updated iteratively. The initial $c_j^0$ is the output of the embedding NN. Any subsequent $c_j^{t+1}$ is the output of a particle-wise NN which takes the invariant density feature vector in the last iteration $\{ \boldsymbol{\rho}_i^t \}$ as the input. The final density feature vector $\{ \boldsymbol{\rho}_i^T \}$ generated by the *T*th interaction block are fed into the corresponding NN module to obtain $\Lambda_{ij}^k(\mathbf{r}_j; \mathbf{r}, \mathbf{R})$ and the coefficients used in the nuclear wavefunction $\psi_n\left(\mathbf{r}, \mathbf{R}\right)$, respectively, as we mentioned above. Note that no cutoff function is presently used so that all nuclei and electrons are included in the nucleus-centered or electron-centered environment.

**Numeric consideration**

In the current approach, the most demanding numerical task is the calculations of the Slater determinants. The calculation of Slater determinants scales as $O(N_e^3)$. Consequently, the overall scaling of SchrödingerNet is approximately $O((N_e+N_n) N_e^3)$, where $O(N_e+N_n)$ is the scaling of the Laplacian calculations. This is similar to the scaling in solving electronic SEs[14], because the electronic DOFs are typically much more than the nuclear DOFs.

Most NN solvers for the electronic SE require $N_s$ repeated training runs to obtain the electronic energy for $N_s$ nuclear configurations when constructing a PES. In contrast, SchrödingerNet can compute the electronic energies for all nuclear configurations within the sampled space in a single run, albeit with an increased sampling space. However, this increase should be modest, as the nuclear DOFs are typically far fewer than the electronic DOFs, especially considering that our designed loss function only requires sampling of the configurational space without integration. Moreover, the calculation of Slater determinants can be significantly accelerated by introducing local approximation and exploiting matrix sparsity, potentially approaching linear scaling[47]. These acceleration schemes will be implemented in future versions.

**RESULTS**

To validate the proposed method, we use SchrödingerNet to solve the SE of a two-electron atom (He), a three-electron atom (Li), and the $H_2^+$ and $H_2$ molecules. The NN structures and training parameters for these systems are detailed in Table I. As mentioned before, a proper symmetry treatment can reduce the search space and allows us to use smaller NNs to obtain similar performance. As shown in Table I, our model uses fewer than 30,000 parameters, demonstrating the efficiency of our ansatz. Notably, this is achieved while fully accounting for the entire quantum system, including both electronic and nuclear DOFs. The incorporation of nuclear DOFs introduces additional complexity by expanding the dimensionality of the system and increasing the range of wavefunction values by several orders of magnitude, as illustrated in Fig. 4, resulting in a more challenging optimization landscape. In comparison, FermiNet[14] utilizes approximately 2,000,000 to 3,000,000 parameters to describe systems with up to several dozen electrons in fixed geometries. Since we focus on the ground state in this work, we adopt the lowest spin configuration in accordance with the Pauli principle. Specifically, for the $H_2$ molecule and He atom, the numbers of spin-up and spin-down electrons are both 1. In the case of Li atom, there are 2 spin-up electrons and 1 spin-down electron.

For the atomic systems, the converged electronic energies are -2.903768 and -7.478097 a.u. for He and Li, respectively. These values are in excellent agreement with the benchmark values of -2.903724 a.u.[48] for He and -7.478067 a.u.[49] for Li.

The atomic systems do not involve nuclear DOFs. To validate the advantage of our method in solving nuclear-electronic SE, we use the $H_2^+$ and $H_2$ molecules as examples. Unlike previous NN-based methods[12-31], which all relied on BOA to solve the electronic SE, SchrödingerNet includes both nuclear and electronic DOFs in the solution. The calculated ground state energy (obtained by the total Hamiltonian acting on total wavefunction, including the nuclear zero-point-energy already) of the $H_2$ molecule is -1.164054 a.u., with a variance of less than $2.4 \times 10^{-5}$ a.u., which agrees very well with results of a benchmark variation-perturbation method[50] (-1.164025 a.u.). This approach also enables us to derive the entire PES by acting the electronic Hamiltonian $\hat{H}_e$ (excluding the nuclear kinetic energy operator from the total Hamiltonian) to the total wavefunction with its nuclear configuration fixed:

$$E(\mathbf{R}) = \sum_{i=1}^{N} \hat{H}_e \psi(\mathbf{r}_i, \mathbf{R}) / \psi(\mathbf{r}_i, \mathbf{R}) / N . \quad (16)$$

In other words, the electronic energy at each nuclear geometry, $E(\mathbf{R})$, is the average of the local energies of MC sampled electronic configurations in the corresponding fixed nuclear coordinates. Fig. 2(a) illustrates the potential energy curve as a function of the distance between two hydrogen atoms. Our results are in excellent agreement with those of an earlier variational calculation with BOA[50].

Importantly, we observed small energy oscillations evidenced by a low variance in the evaluation of electronic energies and ground state energy, as shown in Fig. 3. According to Eq. (17), The low variance makes it possible to efficiently predict the electronic energies and the ground state energies. Indeed, the average of the first 100 step gives a prediction (-1.174449 a.u.), which is already very close to the average of 4000 step (-1.174461 a.u.). In other words, a good estimate can be obtained with a small number of steps.

Using the same strategy, we also solved the SE of the $H_2^+$ molecule and obtained the electronic energy profile of $H_2^+$. As shown in Fig. 2(b), the computational results are in good agreement with the benchmark based on a numerical solution[51], spanning the range of 1.0 ~ 10 a.u.. The calculated ground state energy of $H_2^+$ molecule is -0.597130 a.u., with a variance of less than $1.7 \times 10^{-6}$ a.u., which is in excellent agreement with the free-complementary result (-0.597139 a.u.)[52]. Additionally, the difference in ground-state energies between $H_2$ and $H_2^+$ provides an ionization energy of 0.566924 a.u., which is also in excellent agreement with the experimentally measured value of 0.566889 a.u..[53]

An important observation is that SchrödingerNet is capable of describing the system in the entire nuclear configuration space, including the dissociation, with comparable accuracy in a single training. Bond breaking poses a serious challenge for the couple cluster method due to its single reference nature[6]. In contrast, our NN based approach treats all reasonable nuclear configurations on the same footing, thus the results are uniformly accurate. Indeed, the dissociation energy of $H_2^+$, determined from the difference between the electronic energy at a separation of 10.0 Bohr and the ground state energy, is 0.096982 a.u., which is in good agreement with the experimental value of 0.097412 a.u.[53]. Furthermore, Fig. 4 shows the distribution of configurations in the final training set for the $H_2^+$ molecule. The data points exhibit wide distribution and are not proportional to the square of the wavefunction, validating that our method is not variation-based. Achieving the reported accuracy with a significantly smaller NN size under these challenging conditions highlights the efficiency of our approach.

To provide a visual demonstration of the results, Fig. 5 shows a three-dimensional plot of the total wavefunction of the $H_2^+$ molecule as the function of the electron's x and y coordinates at the internuclear distance of 3.0 a.u. It is clear that SchrödingerNet correctly describes the wavefunction cusp as the electron approaches an H nucleus, which is essential for accurately capturing correlation energy[14]. Furthermore, the significant electronic probability amplitude between the two nuclei is also seen, which is the key for chemical bonding.

We emphasize that SchrödingerNet can be used to compute any molecular properties since the wavefunction is available. For example, it is capable of quantitatively predicting the diagonal Born-Oppenheimer correction (DBOC):

$$E_{\text{DBOC}} = \left\langle \left( -\sum_{I}^{N_n} \frac{1}{2m_I} \nabla_{\mathbf{R}_I}^2 \psi_{el} \right) / \psi_{el} \right\rangle , \tag{17}$$

For $H_2$ in equilibrium configuration, the calculated $E_{DBOC}$ ($5.84 \times 10^{-4}$ a.u.) based on SchrödingerNet using Eq. (17) is in reasonable agreement with that of the full configuration interaction (FCI) calculation ($5.22 \times 10^{-4}$ a.u.)[54]. The small difference is presumably the result of different wavefunction ansatzes and training errors. The nonadiabatic correction is less than 5‰ of the ground state energy, confirming the general validity of the BOA.

**CONCLUSIONS**

To summarize, we propose in this work a simple, accurate, and universal NN solver for the SE that incorporates translational, rotational, parity, and permutational symmetries. The correct treatment of symmetry is not only important for the physically correct properties of the wavefunction, but also helpful to reduce the complexity requirements for the NN structure and the necessary sampling space. SchrödingerNet starts from a physics-inspired ansatz of the wavefunction and introduces a novel loss function aiming to equalize the local energy throughout the system. This ensures that the resulting wavefunction is a good approximation of the eigenstate of a physical system. We emphasize that SchrödingerNet goes beyond the commonly imposed BOA as the wavefunction ansatz requires no separation of the electronic DOFs from those of nuclei. The solution of the full nuclear-electronic SE with SchrödingerNet provides both the energy and wavefunction without quadrature or specific distribution requirements for sampled configurations. Consequently, it allows for the determination of the entire PES in a single training, even at geometries far from the equilibrium. As the wavefunction is available from SchrödingerNet, it is convenient to compute the desired expectation value of an operator. This is illustrated by the calculations of the diagonal Born-Oppenheimer correction. Although it does not rely on the BOA, it can also be used to solve the electronic SE within the BOA, or the nuclear SE with a given PES.

Our method relies on achieving uniformity in local energy across the entire configurational space. This design is well-suited for systems with sparse energy levels but poses challenges for systems with high electronic densities of states. Optimization in such cases becomes difficult as the method struggles to converge to the lowest or desired energy level. One potential solution is to introduce a penalty term into the loss function, such as a weighted loss function ($\mathcal{L} = \sum_i w_1 \left( E_L(\mathbf{r}_i, \mathbf{R}_i) - E \right)^2 + w_1 E$), which simultaneously minimizes local energy variation and the total ground-state energy. However, the effectiveness of this approach requires further validation. In addition, to converge to the correct ground-state energy, the wavefunction must exhibit an accurate behavior in the entire configuration space, including configurations near equilibrium geometries and dissociative asymptotes. Unlike variational methods, this necessitates handling a much larger range of magnitudes in the wavefunction, which may require using a large size NN network. These desirable features make SchrödingerNet a promising ML tool for efficiently solving high-dimensional nuclear-electronic SEs, with a computational scaling of approximately $O((N_e+N_n)N_e^3)$, which is superior to many conventional high-level ab initio methods. We expect it to be useful in tackling many challenging cases such as strongly correlated systems and nonadiabatic effects.

We also note that extensions to excited states are possible, as illustrated in recent work on NN solvers of SE[32].

## DATA AND CODE AVALIABLITY

The SchrödingerNet package are available from https://github.com/zhangylch/SchrodingerNet.

## ACKNOWLEDGEMENTS

This work is supported by US Department of Energy (Grant No. DE-SC0015997 to HG) and by the Strategic Priority Research Program of the Chinese Academy of Sciences (XDB0450101 to BJ), the CAS Project for Young Scientists in Basic Research (YSBR-005 to BJ), the National Natural Science Foundation of China (22325304, 22221003, and 22033007 to BJ). The computation was performed at CARC (Center for Advanced Research Computing) at UNM and the Supercomputing Center of USTC.

## NOTES

The authors declare no competing interests.

## FIGURES & TABLES

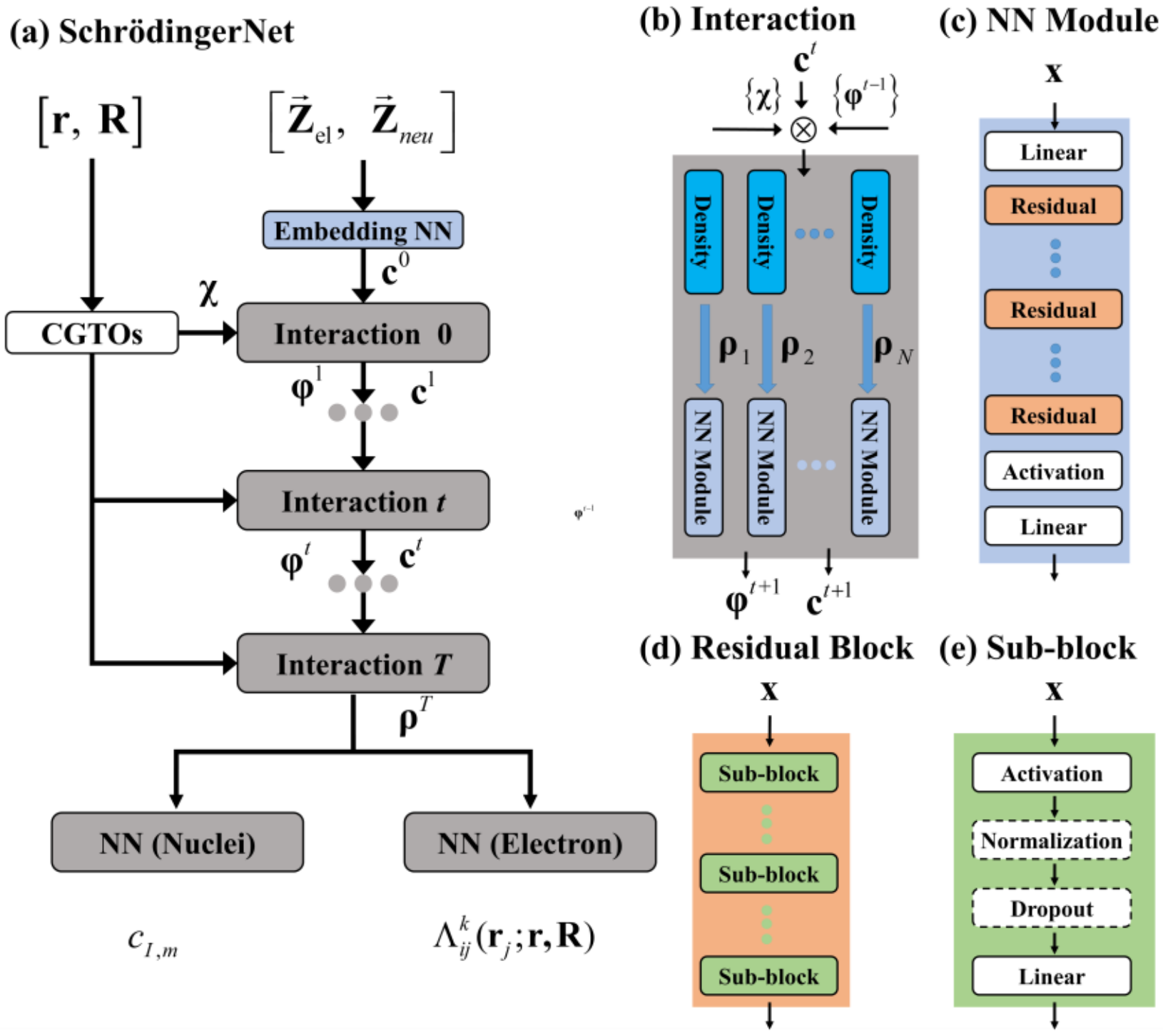

**Fig. 1:** Schematic decomposition and illustration of the SchrödingerNet architecture for solving the full electronic-nuclear SE. The structure of the NN module is inherited from the original recursively embedded atom NN package[47].

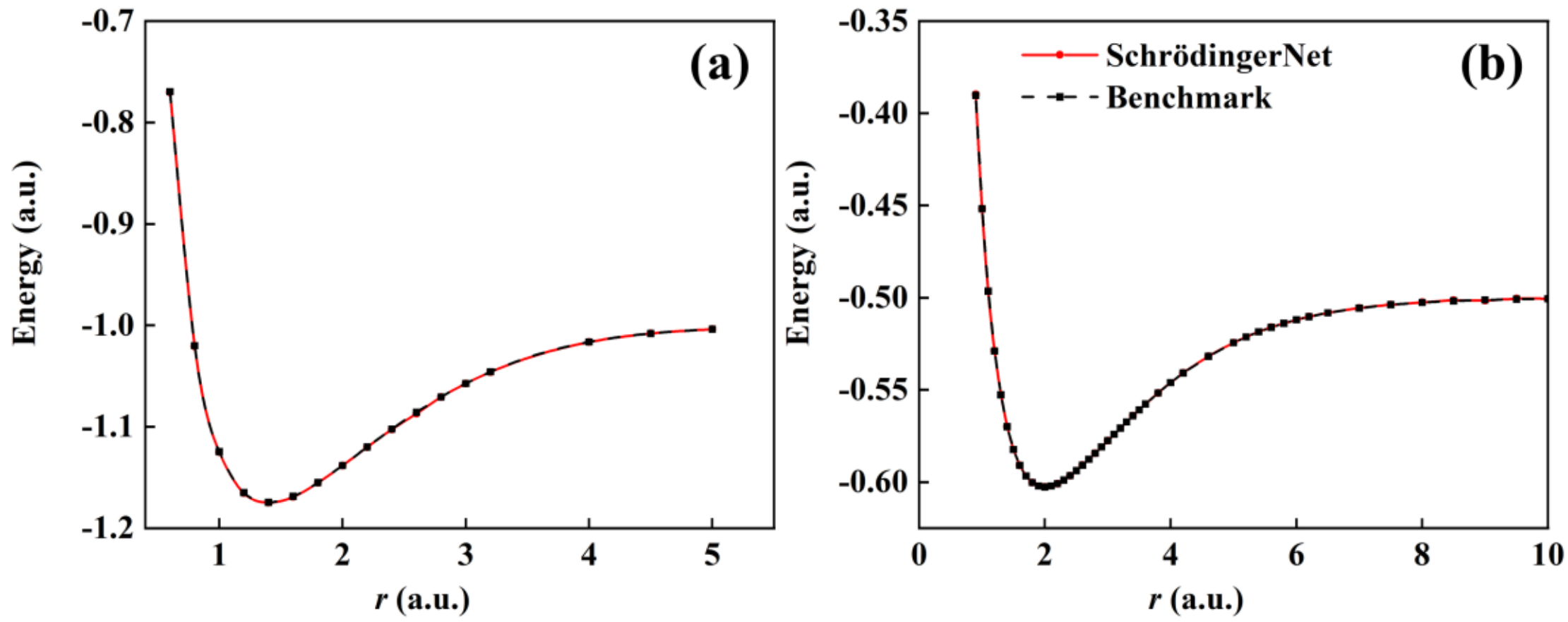

**Fig. 2.** (a) Electronic energies of $H_2$ and (b) $H_2^+$ as a function of the corresponding internuclear distance (r). The results are compared with benchmarks (variational calculation for $H_2$[50] and numerical solution for $H_2^+$[51]).

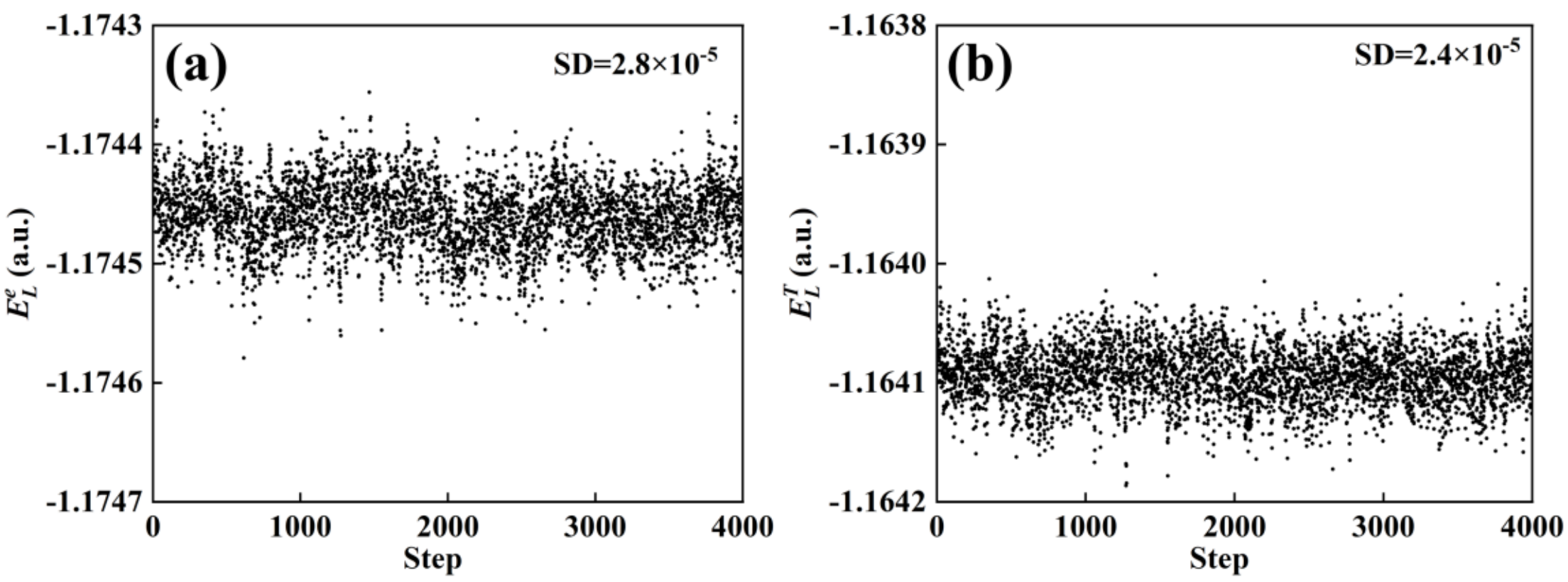

**Fig. 3.** (a) The local electronic energy of $H_2$ (r = 1.4 a.u.) and (b) the local ground state energy of $H_2$ for each MC step (Excluding the first 200 steps from the random initialization of the electronic coordinates). Each point in this figure corresponds to the average of current MC step (4096 configurations). SD represents the standard deviation.

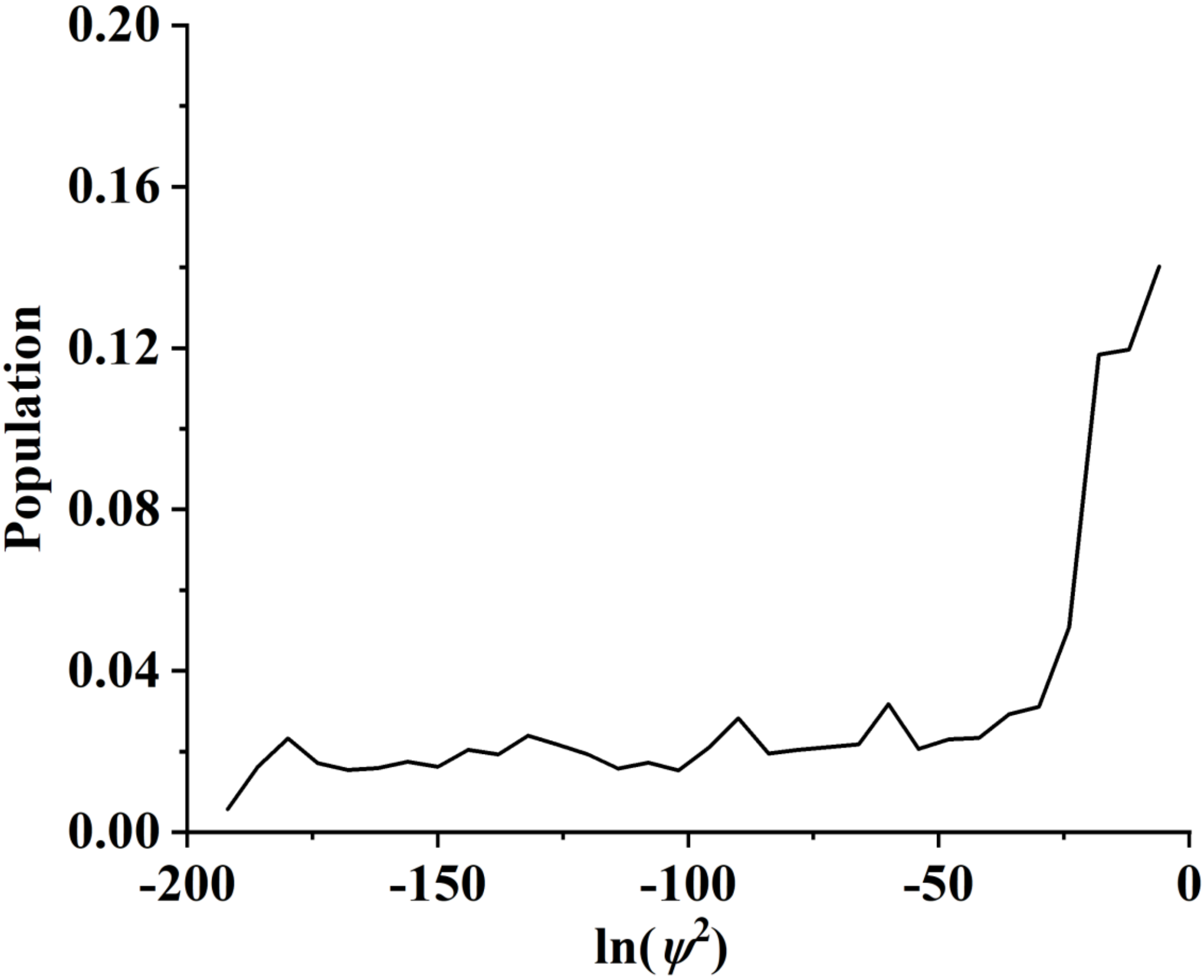

**Fig. 4.** Distribution of the configurations in the final training set for the $H_2^+$ molecule as a function of $\ln\left(\left|\psi\left(\mathbf{r},\ \mathbf{R}\right)\right|^2\right)$

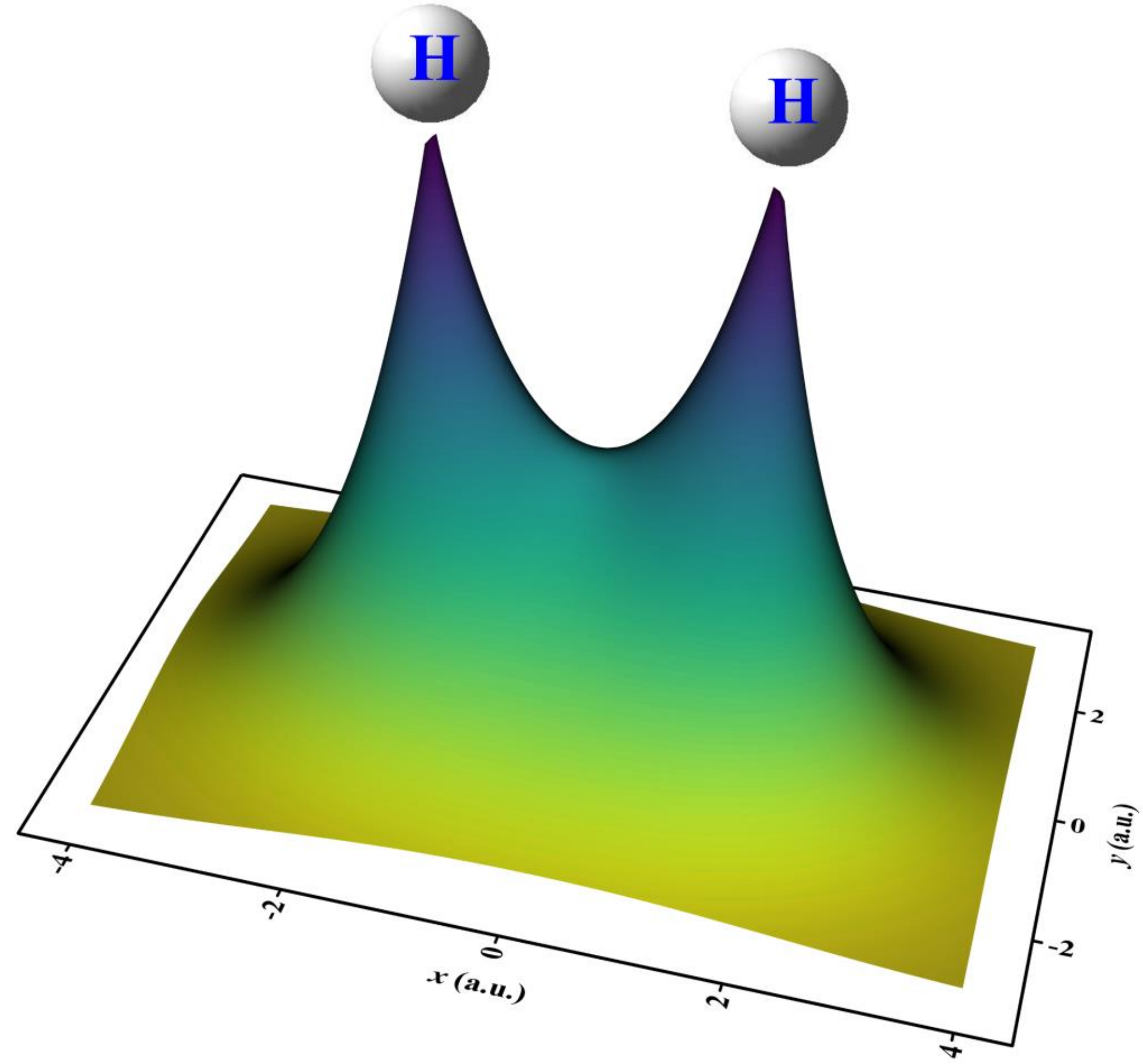

**Fig. 5.** Three-dimensional plot of the total wavefunction of the $H_2^+$ molecule, represented as a function of the electron's x and y coordinates, with the electron's z coordinate fixed at zero. The two hydrogen atoms are positioned at (-1.5, 0, 0) and (1.5, 0, 0) a.u..

Table I: Hyperparameters used in the SchrödingerNet calculations.

| System | He | Li | $H_2$ | $H_2^+$ |
| --- | --- | --- | --- | --- |
| **NN structure**[a] | 64 | 64 | 64 | 64 |
| $N_{block}$[a] | 1 | 1 | 1 | 1 |
| $N_{det}$ | 16 | 16 | 16 | 16 |
| **L** | 1 | 1 | 1 | 1 |
| $N_X$ | 12 | 12 | 12 | 12 |
| **NN structure**[b] | 64×64 | 64×64 | 64×64 | 64×64 |
| **T** | 0 | 1 | 1 | 0 |
| $N_{block}$[b] | 1 | 1 | 1 | 1 |
| $N_{block}$[c] | \ | \ | 1 | 1 |
| $N_{block}$[d] | 1 | 1 | 1 | 1 |

[a], [b], [c] and [d] represent the atomic NN structure for embedding, message passing (iteration 0~T-1), nuclear wavefunction and orbital matrix for the Slater determinant. respectively. $N_{block}$ is the number of residual blocks as shown in Fig. 1(c).

TOC graphic

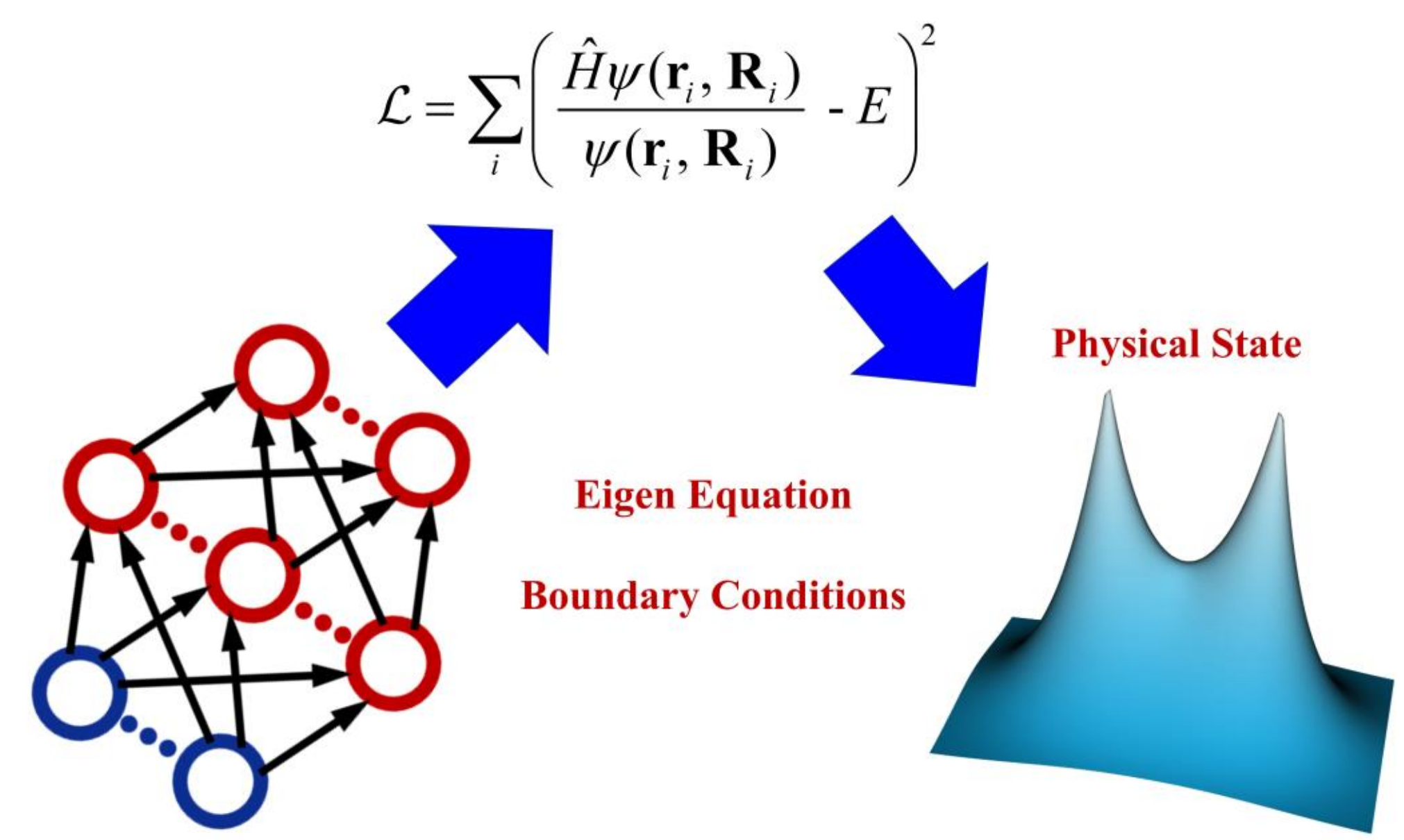